# Four-Fluid Axisymmetric Plasma Equilibrium Model Including Relativistic Electrons and Computational Method and Results

Akio Ishida (石田昭男), Y.-K. Martin Peng (彭元凯)[a] and Wenjun Liu (刘文军)

Hebei Key Laboratory of Compact Fusion, Langfang 065001, China

ENN Science and Technology Development Co., Ltd., Langfang 065001, China

[a] e-mail address: pengyuankai@enn.cn

**Abstract**

A non-relativistic multi-fluid plasma axisymmetric equilibrium model[1] was developed recently to account for the presence of an energetic electron fluid in addition to thermal electron and ion fluids. The equilibrium formulation of a multi-fluid plasma with relativistic energetic electrons is developed and reported in this paper. Relativistic effects in a fluid model approximation can appear in two ways: due to a large macroscopic fluid velocity comparable to the speed of light and a large particle's microscopic random motion which becomes significant if the temperature becomes comparable to or larger than the electron rest mass energy. It is found that the axial component of relativistic generalized angular momentum can be used to describe relativistic axisymmetric equilibrium. The formulation is applied to a four-fluid plasma composed of a relativistic energetic electron fluid, a thermal electron fluid, and fluids of two thermal ion species (e.g. proton and boron ions). The four-fluid density expression which is consistent with the electrostatic potential is obtained and applied in the computation. An example equilibrium approximating a four-fluid plasma recently observed in a solenoid-free ECRH sustained spherical torus plasma[2] is calculated and presented. A second equilibrium that extends the energetic electron temperature of the first example to 679keV is calculated revealing significant relativistic effects.

## I. INTRODUCTION

We adopt in this paper a multi-fluid model to describe macroscopic equilibrium in a solenoid-free ECRH sustained spherical torus plasma, where an energetic electron component can carry a dominant fraction of the plasma current.[3, 4] Hard X-ray spectra measured on QUEST showed that temperature of the bulk of the energetic electron component can be 50 keV in an energy range up to 600 keV.[4] Using ECRH power from two frequencies, hard X-ray temperatures of the energetic electron component as high as 500 keV were measured.[5] Such electrons were collisionless and strongly decoupled from the low-temperature plasma ions and electrons that were also present. It is therefore appropriate, as a first approximation, to describe these plasmas using a multi-fluid equilibrium model that includes a relativistic electron component.

If the velocity of a fluid within a small volume becomes comparable to the speed of light, we must take into account of the well-known relativistic effect.[6] With fluid motion, this effect changes the number density of the fluid element due to the Lorentz contraction.[7] A macroscopic fluid element contains constituent particles. A second relativistic effect is therefore due to the random motion of these particles, as the fluid temperature becomes comparable with or more than their rest-mass energy.[7,8] To account for the above-mentioned effects, relativistic magnetofluid model is adopted to describe the energetic electron component, similar to what has been developed and used to calculate the equilibrium magnetic field structure in astrophysics[9-12]. Recently we developed an axisymmetric multi-fluid equilibrium model assuming a non-relativistic energetic electron fluid.[1] In this paper, we describe an equilibrium formulation of a relativistic multi-fluid plasma and its application to a four-fluid plasma.

Multi-fluid axisymmetric equilibrium formulation was developed earlier for collisional plasmas[13] and then for weakly-collisional plasmas.[14] The latter model was applied[15] to the NSTX experiment at the Princeton Plasma Physics Laboratory. The reconstructed equilibrium profiles were consistent with the TRANSP[16] calculations, revealing that the ion diamagnetic drift was related to the radial electric field in the edge region (Figures 1-6 of Ref. 15). More recently a multi-ion axisymmetric equilibrium model was developed and applied to D-$^3$He and p-$^{11}$B plasmas in the FRC configuration[17], where the poloidal plasma flow is absent. We adopt in the present paper the formulation developed in Ref.14, where a poloidal plasma flow is included, and extend the model to include a relativistic electron fluid.

Although a toroidal equilibria of a relativistic electron beam plasma was modeled in the large-aspect ratio limit,[18] our present model deals with a multi-fluid plasma with a relativistic component that is valid in low-aspect ratio. As far as we know, the formulation of this type is original.

This paper is organized as follows. Section II describes the equations of relativistic magnetofluid dynamics in three dimensions. It shows that the generalized momentum of the relativistic magnetofluid plays an integral part in the model. Section III describes the axisymmetric formulation for the relativistic magneto-multi-fluid equilibrium. The computational method for a four-fluid plasma model is described in Section IV. Two example equilibria close to measurements obtained on EXL-50[2] are provided in Section V. Further application of this equilibrium model to the analysis of today's toroidal plasmas containing a relativistic electron component is discussed in Section VI. Appendix A describes in detail the thermodynamic property of an ideal relativistic fluid.

Appendix B describes the surface profile functions used in the computation. Appendix C describes the formulation used to verify that the radial force balance of each fluid species is satisfied.

## II. EQUATIONS OF RELATIVISTIC MAGNETOFLUID DYNAMICS

Although we study a multi-fluid plasma model in this paper, we discuss in this section the relativistic forms of the continuity equation and the equation of motion for any single species, neglecting subscript which depicts species concerned for simplicity of notation. The continuity equation is given by

$$\frac{\partial \gamma n}{\partial t} + \nabla \cdot (\gamma n \boldsymbol{u}) = 0. \tag{1}$$

Here $n$ is the density in the proper frame in which the fluid volume element concerned is at rest, $\boldsymbol{u}$ is the fluid velocity and $\gamma \equiv (1 - u^2/c^2)^{-1/2}$ is the Lorentz factor arising in derivations of the Lorentz transformations.[6] Note that $\gamma n$ is the density in the laboratory frame in which the fluid volume element concerned moves with $\boldsymbol{u}$ and the factor $\gamma$.[7] Suppose that the density is defined as $n = \Delta N / \Delta V$ in the proper frame where $\Delta V$ is the volume and $\Delta N$ is number of particles contained in the volume $\Delta V$. In the laboratory frame where this volume element moves with velocity $\boldsymbol{u}$, the volume $\Delta V$ shrinks to $\Delta V / \gamma$ by Lorentz contraction. Since $\Delta N$ is not changed, the density in the laboratory frame becomes $\gamma n$.

The equation of fluid motion is given by

$$\frac{\partial}{\partial t}\left(mn\gamma^2 g_{ep}(T^*)\boldsymbol{u}\right) + \frac{\partial}{\partial x_j}\left(mn\gamma^2 g_{ep}(T^*)\boldsymbol{u}u_j\right) + \nabla p = q\gamma n \boldsymbol{E} + q\gamma n \boldsymbol{u} \times \boldsymbol{B}, \tag{2}$$

where $m$ and $q$ are the mass and charge of a particle consisting of the species fluid concerned, $p$ is the pressure, $\boldsymbol{E}$ and $\boldsymbol{B}$ are the electric and magnetic fields, respectively.[6] Here it is understood to take sum of j=1, 2 and 3 for the repeated j where $x_1 = x$, $x_2 = y$, and $x_3 = z$. The quantity $g_{ep}(T^*)$ is defined by Eq. (3), which is the relativistic energy density ratio over that of the fluid rest-mass energy density.

$$g_{ep}(T^*) \equiv \frac{\epsilon + p}{mnc^2} = \frac{K_3(1/T^*)}{K_2(1/T^*)}, \tag{3}$$

where $\epsilon$ is the energy density of the fluid which contains rest mass energy, $T^* \equiv T/mc^2$,[8] $T$ is the temperature of the fluid in energy unit and $K_n(z)$ is the 2$^{nd}$ kind modified Bessel function defined by

$$K_n(z) = \frac{\sqrt{\pi}\left(\frac{z}{2}\right)^n}{\Gamma\left(n+\frac{1}{2}\right)} \int_1^\infty dt\, e^{-zt}(t^2 - 1)^{n-\frac{1}{2}}. \tag{4}$$

Derivation of Eq. (3) is shown in Appendix A. Note that $\epsilon + p$ is the enthalpy density or the heat function per unit volume.[7] The unfamiliar quantity, $mn\gamma^2 g_{ep}\boldsymbol{u}$, should be understood as $\gamma n \times$ $m\gamma \boldsymbol{u} \times g_{ep}$ where $\gamma n$ is density, $m\gamma \boldsymbol{u}$ is momentum due to macroscopic motion in the laboratory frame and $g_{ep}$ represents relativistic effect due to random motion of particles contained in a fluid element concerned. Since $\gamma^2 g_{ep}$ becomes unity in the non-relativistic limit (see Appendix A), Eq. (2) is convenient to know the non-relativistic limit.

The relativistic magnetofluid model has been used to describe relativistic flows from rapid rotators[9], relativistic axisymmetric jet equilibrium[10], relativistic large-scale magnetic reconnection[11] and twisted torus magnetic configuration of magnetars[12].

In the four-dimensional form of relativistic equations, the cgs-Gauss unit is convenient because $\boldsymbol{E}$ and $\boldsymbol{B}$ can transfer to each other by Lorentz transformation. As equations (1) and (2) are the three-dimensional form, we adopt the SI unit. Using the vector and scaler potentials, $\boldsymbol{E}$ and $\boldsymbol{B}$ are expressed as

$$\boldsymbol{E} = -\frac{\partial \boldsymbol{A}}{\partial t} - \nabla V_E \quad \text{and} \quad \boldsymbol{B} = \nabla \times \boldsymbol{A}, \tag{5}$$

where $V_E$ is the electrostatic potential and $\boldsymbol{A}$ is the magnetic vector potential. Using the continuity equation and the above equation in Eq. (2) and multiplying $1/n$, the resultant equation is given by

$$\gamma \frac{\partial}{\partial t}(m\gamma g_{ep}\boldsymbol{u} + q\boldsymbol{A}) + \gamma(\boldsymbol{u}\cdot\nabla)(m\gamma g_{ep}\boldsymbol{u}) + \nabla T + T\nabla \ln n + q\gamma \nabla V_E = \gamma \boldsymbol{u} \times (\nabla \times q\boldsymbol{A}), \tag{6}$$

where $p = nT$ is used. See Appendix A. Rewriting the right side of Eq. (6) as,

$\gamma \boldsymbol{u} \times (\nabla \times q\boldsymbol{A}) = \gamma \boldsymbol{u} \times (\nabla \times (m\gamma g_{ep}\boldsymbol{u} + q\boldsymbol{A})) - \gamma \boldsymbol{u} \times (\nabla \times (m\gamma g_{ep}\boldsymbol{u}))$,

we have

$$\gamma \frac{\partial}{\partial t}\boldsymbol{P} + \nabla T + T\nabla \ln n + q\gamma \nabla V_E + \gamma(\boldsymbol{u}\cdot\nabla)(m\gamma g_{ep}\boldsymbol{u}) + \gamma \boldsymbol{u} \times (\nabla \times (m\gamma g_{ep}\boldsymbol{u})) = \gamma \boldsymbol{u} \times (\nabla \times \boldsymbol{P}), \tag{7}$$

where $\boldsymbol{P} \equiv m\gamma g_{ep}\boldsymbol{u} + q\boldsymbol{A}$ is the generalized momentum. In order to rewrite the last two terms in the L.H.S. of Eq. (7), we use the following vector identity,

$\nabla(\boldsymbol{M}\cdot\boldsymbol{N}) = \boldsymbol{M} \times (\nabla \times \boldsymbol{N}) + \boldsymbol{N} \times (\nabla \times \boldsymbol{M}) + (\boldsymbol{M}\cdot\nabla)\boldsymbol{N} + (\boldsymbol{N}\cdot\nabla)\boldsymbol{M}$

where $\boldsymbol{M}$ and $\boldsymbol{N}$ are arbitrary vectors. First take $\boldsymbol{M} = \boldsymbol{u}$, $\boldsymbol{N} = \gamma g_{ep}\boldsymbol{u}$ and second take $\boldsymbol{M} = \boldsymbol{N} = \boldsymbol{u}$. Then, the last two terms in the L.H.S. of Eq. (7) can be written as

$$\gamma(\boldsymbol{u}\cdot\nabla)(m\gamma g_{ep}\boldsymbol{u}) + \gamma \boldsymbol{u} \times (\nabla \times (m\gamma g_{ep}\boldsymbol{u})) = \gamma u \nabla(m\gamma u g_{ep}), \tag{8}$$

where $u$ is the magnitude of the velocity $\boldsymbol{u}$. Substituting Eq. (8) into Eq. (7) leads to

$$\gamma \frac{\partial}{\partial t}\boldsymbol{P} + m\gamma u \nabla(\gamma u g_{ep}) + \nabla T + T\nabla \ln n + q\gamma \nabla V_E = \gamma \boldsymbol{u} \times (\nabla \times \boldsymbol{P}). \tag{9}$$

This is the required form of equation of motion for a "fluid element". As the "fluid element" is macroscopic, it contains many microscopic particles. As a result, Eq. (9) describing the behavior of generalized momentum $\boldsymbol{P}$ accounts for the effects of pressure gradient, electric field, relativistic macroscopic velocity through $\gamma$, and relativistic velocity of microscopic random motion through $g_{ep}(T^*)$. For axisymmetric plasma, it is easily verified that Eq. (9) conserves the z-component of the generalized angular momentum, $\hat{z} \cdot (\boldsymbol{R} \times \boldsymbol{P}) = RP_\phi$, i.e. Eq. (10) is satisfied.

$$\frac{\partial RP_\phi}{\partial t} + \boldsymbol{u} \cdot \boldsymbol{\nabla}(RP_\phi) = 0, \tag{10a}$$

where $RP_\phi = m\gamma g_{ep} R u_\phi + qRA_\phi$. (10b)

This quantity plays a key role in axisymmetric equilibrium formulation.

## III. AXISYMMETRIC EQUILIBIUM FORMULATION

For each species fluid, we adopt following two equations from Eq. (1) and Eq. (9).

$$\nabla \cdot (\gamma n \boldsymbol{u}) = 0, \tag{11}$$

$$m\gamma \boldsymbol{u} \boldsymbol{\nabla}(\gamma u g_{ep}) + \boldsymbol{\nabla} T + T\boldsymbol{\nabla}\ln n + q\gamma \boldsymbol{\nabla} V_E = q\gamma \boldsymbol{u} \times \boldsymbol{\Omega}, \tag{12}$$

where $\boldsymbol{\Omega} \equiv q^{-1} \boldsymbol{\nabla} \times \boldsymbol{P} = \boldsymbol{B} + \boldsymbol{\nabla} \times (q^{-1} m\gamma g_{ep} \boldsymbol{u})$. (13)

In this paper we call $\boldsymbol{\Omega}$ the modified magnetic field. Maxwell's equations govern the fields

$$\boldsymbol{\nabla} \times \boldsymbol{B} = \mu_0 \sum_\alpha \boldsymbol{j}_\alpha \text{ where } \boldsymbol{j}_\alpha = q_\alpha \gamma_\alpha n_\alpha \boldsymbol{u}_\alpha, \tag{14}$$

$$\boldsymbol{\nabla} \cdot \boldsymbol{B} = 0. \tag{15}$$

We adopt the charge neutrality condition instead of Poisson's equation,

$$\sum_\alpha q_\alpha \gamma_\alpha n_\alpha = 0. \tag{16}$$

Hereafter we adopt the right-hand cylindrical coordinates $(R, \phi, Z)$. As the magnetic field $\boldsymbol{B}$, the modified magnetic field $\boldsymbol{\Omega}$ and the density flux $\gamma n \boldsymbol{u}$ are divergence-free, these can be expressed using flux or stream functions $\psi, Y$ and $\Phi$.

$$\boldsymbol{B} = \boldsymbol{\nabla}\psi(R,Z) \times \boldsymbol{\nabla}\phi + RB_\phi \boldsymbol{\nabla}\phi, \tag{17}$$

$$\boldsymbol{\Omega} = \boldsymbol{\nabla} Y(R,Z) \times \boldsymbol{\nabla}\phi + R\Omega_\phi \boldsymbol{\nabla}\phi, \tag{18}$$

$$n\gamma \boldsymbol{u} = \boldsymbol{\nabla}\Phi(R,Z) \times \boldsymbol{\nabla}\phi + Rn\gamma u_\phi \boldsymbol{\nabla}\phi. \tag{19}$$

Note that since $\boldsymbol{\Omega}$ is defined by Eq. (13), the following relation must be satisfied,

$$Y = \psi + \frac{m}{q} \gamma g_{ep} R u_\phi, \tag{20}$$

$$R\Omega_\phi = RB_\phi - R^2 \boldsymbol{\nabla} \cdot \left(\frac{mg_{ep}}{qnR^2} \boldsymbol{\nabla}\Phi\right). \tag{21}$$

### A. Force balance in directions of $\boldsymbol{\nabla}\phi, \boldsymbol{\Omega}$ and $\boldsymbol{\nabla} Y$

Since in axisymmetric equilibrium, the L.H.S. of Eq. (12) has no $\phi$-component, following relation must be satisfied,

$$\boldsymbol{\nabla}\phi \cdot (\boldsymbol{u} \times \boldsymbol{\Omega}) = 0. \tag{22}$$

This gives $\nabla\phi \cdot (\nabla\Phi(R,Z) \times \nabla Y(R,Z)) = 0$ for arbitrary R and Z. The above is satisfied when

$$\Phi = \Phi(Y(R,Z)), \tag{23}$$

i.e. the function $\Phi$ is arbitrary function of $Y(R,Z)$.

We adopt following assumption,

$$\mathbf{\Omega} \cdot \nabla T = 0. \tag{24}$$

Validity of this assumption is discussed in Ref. 14. This is natural generalization of $\mathbf{B} \cdot \nabla T = 0$ for flowing fluid. Key points are as follows: 1) Species poloidal flow velocity is much smaller than the same species thermal velocity, 2) For axisymmetric flowing plasma, $\mathbf{\Omega} \equiv q^{-1}\nabla \times \mathbf{P}$ is more fundamental than $\mathbf{B}$, and 3) Temperature anisotropy is negligible for TST-2 ohmic plasmas.[19, 20] This assumption is used successfully for non-relativistic two-fluid equilibrium[15] and three-fluid equilibrium[1]. The above assumption, Eq. (24), leads to

$$T = T(Y(R,Z)) \quad \text{and} \quad g_{ep} = g_{ep}(T(Y(R,Z))/mc^2). \tag{25}$$

i.e. the function $T$ is arbitrary function of $Y(R,Z)$.

Next, consider the force balance in $\mathbf{\Omega}$-direction. This is given by

$$\nabla\phi \cdot (\{m\gamma u \nabla(\gamma u g_{ep}) + \nabla T + T\nabla \ln n + q\gamma \nabla V_E\} \times \nabla Y) = 0. \tag{26}$$

Using Eq. (25), the above can be written as

$$\nabla\phi \cdot (\nabla\{\tfrac{1}{2}m(\gamma u)^2 g_{ep} + T(1 + \ln n) + qV_E\} \times \nabla Y) + \nabla\phi \cdot \{q(\gamma-1)\nabla V_E \times \nabla Y\} = 0. \tag{27}$$

Hereafter we neglect the second term in Eq. (27). This is equivalent to put $\gamma = 1$ in the electric force term in Eq. (12). As the Lorentz force is much larger that the electric force in general, this neglect is not harmful. Define the function $F$ as

$$F \equiv \tfrac{1}{2}m(\gamma u)^2 g_{ep} + T(1 + \ln n) + qV_E. \tag{28}$$

Within this approximation, Eq. (27) can be satisfied for arbitrary R and Z when

$$F = F(Y(R,Z)), \tag{29}$$

i.e. $F$ must be arbitrary function of $Y(R,Z)$. Finally, consider the force balance in the $\nabla Y$ direction. Using Eq. (29), the force balance equation (12) can be written as within the approximation described above

$$\left(\frac{dF}{dY} - \frac{dT}{dY}\ln n + \frac{1}{2}m(\gamma u)^2 \frac{dg_{ep}}{dT}\frac{dT}{dY}\right)\nabla Y = \frac{q}{R}\left(\gamma u_\phi - \frac{1}{n}\frac{d\Phi}{dY}\Omega_\phi\right)\nabla Y. \tag{30}$$

While the Grad-Shafranov model has two arbitrary functions of magnetic flux $\psi$, in the present model each species fluid has three arbitrary functions $F(Y)$, $T(Y)$ and $\Phi(Y)$ where $Y$ is given by Eq. (20). In this paper we call these functions the profile functions. Note also that $Y = RP_\phi/q$ which is the z-component of the generalized angular momentum divided by charge $q$. See Eq. (10b).

Solving Eq. (30) for the toroidal flow velocity, we have for each species $\alpha$

$$\gamma_\alpha u_{\alpha\phi} = q_\alpha^{-1} R \left( F_\alpha' - T_\alpha' \ln n_\alpha + \tfrac{1}{2} m_\alpha (\gamma_\alpha u_\alpha)^2 \frac{dg_{ep}}{dT_\alpha} T_\alpha' \right) + \frac{1}{n_\alpha} \Phi_\alpha' \Omega_{\alpha\phi} \tag{31}$$

where the prime designates derivative with respect to $Y_\alpha$. The toroidal current density of each species is given by

$$j_{\alpha\phi} = q_\alpha n_\alpha \gamma_\alpha u_{\alpha\phi} = n_\alpha R \left( F_\alpha' - T_\alpha' \ln n_\alpha + \tfrac{1}{2} m_\alpha (\gamma_\alpha u_\alpha)^2 \frac{dg_{ep}}{dT_\alpha} T_\alpha' \right) + q_\alpha \Phi_\alpha' \Omega_{\alpha\phi}. \tag{32}$$

## B. Ampere's law

Using Eq. (17),

$$\nabla \times \boldsymbol{B} = \nabla(RB_\phi) \times \nabla\phi - R \left\{ \frac{\partial}{\partial R}\left(\frac{1}{R}\frac{\partial \psi}{\partial R}\right) + \frac{1}{R}\frac{\partial^2 \psi}{\partial Z^2} \right\} \nabla\phi. \tag{33}$$

The toroidal component of Ampere's law is written as

$$\frac{\partial}{\partial R}\left(\frac{1}{R}\frac{\partial \psi}{\partial R}\right) + \frac{1}{R}\frac{\partial^2 \psi}{\partial Z^2} = -\mu_0 j_\phi \tag{34}$$

where $j_\phi = \sum_\alpha j_{\alpha\phi}$. Using Eq. (19), the poloidal component of Ampere's law is written as

$$RB_\phi = \sum_\alpha \mu_0 q_\alpha \Phi_\alpha. \tag{35}$$

Explicit expression of the charge neutrality condition becomes complex as number of fluids constituting a plasma increases. We will discuss its application in the next section.

## IV. COMPUTAIONAL METHOD FOR FOUR-FLUID PLASMA MODEL

A plasma considered here consists of two ion species (proton and an arbitrary dominant impurity ion, e.g., boron) and two electron species (low temperature and high density electron component and high temperature, high speed, and low density electron component). The subscript $\alpha = p, im, el, eh$ represent the proton fluid, the impurity ion fluid, the low temperature and high density electron fluid and the high temperature, high speed and low density electron fluid, respectively.

As for the eh-fluid, we adopt the relativistic form of equations derived in section III while as for the other three fluids, we adopt non-relativistic form of equations putting $\gamma_\alpha = 1$, $g_{ep\alpha} = 1$ and $\frac{dg_{ep\alpha}}{dT_\alpha} = 0$ for $\alpha = p, im, el$ because the non-relativistic limit takes $c \to \infty$, i.e. speed of light is infinity.

### A. Dimensionless form of equilibrium equations

As dimensionless form is convenient in numerical computation we adopt dimensionless variables hereafter. The primary scales are (1) $L_{ref}$ a reference length of a plasma, (2) $I_{ref}$ a reference plasma current and (3) $n_{ref}$ a reference density. These lead to derived scales for the magnetic field $B_{ref} \equiv \frac{\mu_0 I_{ref}}{L_{ref}}$, velocities $u_{ref} \equiv B_{ref}/(\mu_0 m_p n_{ref})^{1/2}$, temperatures $T_{ref} \equiv m_p u_{ref}^2$, pressures $p_{ref} \equiv n_{ref} T_{ref}$, scalar potential $V_{Eref} \equiv T_{ref}/e$, vector potential $A_{ref} \equiv B_{ref} L_{ref}$, poloidal magnetic flux $\psi_{ref} \equiv B_{ref} L_{ref}^2$, current density $j_{ref} \equiv B_{ref}/(\mu_0 L_{ref})$, poloidal flow surface function $\Phi_{ref} \equiv n_{ref} u_{ref} L_{ref}^2$, the function $F_{ref} = T_{ref}$, and electric field $E_{ref} \equiv u_{ref} B_{ref}$. Introduce parameter $\varepsilon \equiv l_{pR}/L_{ref}$ where $l_{pR} \equiv \frac{c}{\sqrt{e^2 n_{ref}/\varepsilon_0 m_p}}$ is the proton inertial length for the reference density $n_{ref}$.

For simplicity we neglect symbol which depicts dimensionless variables unless otherwise noted. The followings are dimensionless equations.

$$F_p = \frac{1}{2} u_p^2 + T_p(1 + \ln n_p) + V_E = F_p(Y_p) \tag{36a}$$

$$F_{im} = \frac{1}{2} \frac{m_{im}}{m_p} u_{im}^2 + T_{im}(1 + \ln n_{im}) + Z_{im} V_E = F_{im}(Y_{im}) \tag{36b}$$

$$F_{el} = \frac{1}{2} \frac{m_e}{m_p} u_{el}^2 + T_{el}(1 + \ln n_{el}) - V_E = F_{el}(Y_{el}) \tag{36c}$$

$$F_{eh} = \frac{1}{2} \frac{m_e}{m_p} u_{eh}^2 \gamma_{eh}(u_{eh})^2 g_{ep}(T_{eh}^*) + T_{eh}(1 + \ln n_{eh}) - V_E = F_{eh}(Y_{eh}) \tag{36d}$$

where $\gamma_{eh}(u_{eh}) = (1 - u_{eh}^2/\bar{c}^2)^{-1/2}$, $T_{eh}^* = T_{eh}/(\frac{m_e}{m_p}\bar{c}^2)$, $\bar{c} \equiv c/u_{ref}$ and

$$Y_p = \psi + \varepsilon R u_{p\phi} \tag{37a}$$
$$Y_{im} = \psi + \varepsilon (m_{im}/m_p Z_{im}) R u_{im\phi} \tag{37b}$$
$$Y_{el} = \psi - \varepsilon (m_e/m_p) R u_{el\phi} \tag{37c}$$
$$Y_{eh} = \psi - \varepsilon (m_e/m_p) \gamma_{eh}(u_{eh}) g_{ep}(T_{eh}^*) R u_{eh\phi}. \tag{37d}$$

In order to select the external toroidal field easily, we introduce new function for poloidal flow defined as

$$\Phi_\alpha(Y_\alpha) = -\varepsilon K_\alpha(Y_\alpha(R,Z)) \quad \text{for } \alpha = p, im, el, eh. \tag{38}$$

The toroidal flow velocities are written as

$$u_{p\phi} = \varepsilon R(F_p' - T_p' \ln n_p) - \frac{\varepsilon}{n_p} K_p' \Omega_{p\phi} \tag{39a}$$

$$u_{im\phi} = \frac{\varepsilon}{Z_{im}} R(F_{im}' - T_{im}' \ln n_{im}) - \frac{\varepsilon}{n_{im}} K_{im}' \Omega_{im\phi} \tag{39b}$$

$$u_{el\phi} = -\varepsilon R(F_{el}' - T_{el}' \ln n_{el}) - \frac{\varepsilon}{n_{el}} K_{el}' \Omega_{el\phi} \tag{39c}$$

$$\gamma_{eh} u_{eh\phi} = -\varepsilon R \left( F_{eh}' - T_{eh}' \ln n_{eh} + \frac{1}{2} \frac{m_e}{m_p} (\gamma_{eh}(u_{eh}) u_{eh})^2 \frac{d g_{ep}(T_{eh}^*)}{d T_{eh}} T_{eh}' \right) - \frac{\varepsilon}{n_{eh}} K_{eh}' \Omega_{eh\phi} \tag{39d}$$

where

$$\Omega_{\alpha\phi} = B_\phi + \varepsilon^2 \frac{e}{q_\alpha} \frac{m_\alpha}{m_p} R \nabla \cdot \left( \frac{K'_\alpha}{n_\alpha R^2} \nabla Y_\alpha \right) \quad \text{for } \alpha = p, im, el \tag{40a}$$

$$\Omega_{eh\phi} = B_\phi - \varepsilon^2 \frac{m_e}{m_p} R \nabla \cdot \left( \frac{g_{epeh} K'_{eh}}{n_{eh} R^2} \nabla Y_{eh} \right). \tag{40b}$$

The toroidal current densities are written as

$$j_{p\phi} = \varepsilon^{-1} n_p u_{p\phi} \tag{41a}$$

$$j_{im\phi} = \varepsilon^{-1} Z_{im} n_{im} u_{im\phi} \tag{41b}$$

$$j_{el\phi} = -\varepsilon^{-1} n_{el} u_{el\phi} \tag{41c}$$

$$j_{eh\phi} = -\varepsilon^{-1} n_{eh} \gamma_{eh} u_{eh\phi}. \tag{41d}$$

The toroidal component of Ampere's law is written as

$$R \frac{\partial}{\partial R} \left( \frac{1}{R} \frac{\partial \psi}{\partial R} \right) + \frac{\partial^2 \psi}{\partial Z^2} = -R j_\phi \tag{42}$$

where $j_\phi = j_{p\phi} + j_{im\phi} + j_{el\phi} + j_{eh\phi}.$ \hfill (43)

The poloidal component of Ampere's law is written as

$$R B_\phi = (K_{el} + K_{eh} - K_p - Z_{im} K_{im}). \tag{44}$$

Solving Eq. (36) for the electrostatic potential $V_E$, we have

$$V_E = \left( \widetilde{F}_p - T_p (1 + \ln n_p) \right) \tag{45a}$$

$$V_E = \frac{1}{Z_{im}} \left( \widetilde{F}_{im} - T_{im} (1 + \ln n_{im}) \right) \tag{45b}$$

$$V_E = -\left( \widetilde{F}_{el} - T_{el} (1 + \ln n_{el}) \right) \tag{45c}$$

$$V_E = -\left( \widetilde{F}_{eh} - T_{eh} (1 + \ln n_{eh}) \right) \tag{45d}$$

where

$$\widetilde{F}_p \equiv F_p - \frac{1}{2} u_p^2 \tag{46a}$$

$$\widetilde{F}_{im} \equiv F_{im} - \frac{1}{2} \frac{m_{im}}{m_p} u_{im}^2 \tag{46b}$$

$$\widetilde{F}_{el} \equiv F_{el} - \frac{1}{2} \frac{m_e}{m_p} u_{el}^2 \tag{46c}$$

$$\widetilde{F}_{eh} \equiv F_{eh} - \frac{1}{2} \frac{m_e}{m_p} u_{eh}^2 \gamma_{eh}(u_{eh})^2 g_{ep}(T^*_{eh}). \tag{46d}$$

## B. Equations of densities consistent with electrostatic potential

Eliminating $V_E$ from Eq.(45a) and Eq. (45c),

$$\widetilde{F}_p + \widetilde{F}_{el} = (T_p + T_{el})(1 + \ln n_p) + T_{el} \ln \frac{n_{el}}{n_p}. \tag{47a}$$

Likewise, from Eq. (45a) and Eq. (45d),

$$\widetilde{F}_p + \widetilde{F}_{eh} = (T_p + T_{eh})(1 + \ln n_p) + T_{eh}\ln\frac{n_{eh}}{n_p}. \tag{47b}$$

From Eq. (45a) and Eq. (45b),

$$\widetilde{F}_p - \frac{\widetilde{F}_{im}}{Z_{im}} = \left(T_p - \frac{T_{im}}{Z_{im}}\right)(1 + \ln n_p) - \frac{T_{im}}{Z_{im}}\ln\frac{n_{im}}{n_p}. \tag{47c}$$

Eliminating $(1 + \ln n_p)$ from Eq. (47a,b), we have

$$\frac{\widetilde{F}_p + \widetilde{F}_{el} - T_{el}\ln\frac{n_{el}}{n_p}}{T_p + T_{el}} = \frac{\widetilde{F}_p + \widetilde{F}_{eh} - T_{eh}\ln\frac{n_{eh}}{n_p}}{T_p + T_{eh}}. \tag{48a}$$

Likewise, eliminating $(1 + \ln n_p)$ from Eq. (47b,c), we have

$$\frac{\widetilde{F}_p + \widetilde{F}_{eh} - T_{eh}\ln\frac{n_{eh}}{n_p}}{T_p + T_{eh}} = \frac{\widetilde{F}_p - \frac{\widetilde{F}_{im}}{Z_{im}} + \frac{T_{im}}{Z_{im}}\ln\frac{n_{im}}{n_p}}{T_p - \frac{T_{im}}{Z_{im}}}. \tag{48b}$$

Solving Eq. (48b) for $n_{im}/n_p$, we have

$$\frac{n_{im}}{n_p} = \exp\left\{\frac{\frac{\widetilde{F}_{im}}{Z_{im}}}{\frac{T_{im}}{Z_{im}}} + \frac{\left(T_p - \frac{T_{im}}{Z_{im}}\right)\widetilde{F}_{eh} - \left(T_{eh} + \frac{T_{im}}{Z_{im}}\right)\widetilde{F}_p}{\frac{T_{im}}{Z_{im}}(T_p + T_{eh})} - \frac{T_p - \frac{T_{im}}{Z_{im}}}{\frac{T_{im}}{Z_{im}}}\frac{T_{eh}}{(T_p + T_{eh})}\ln\frac{n_{eh}}{n_p}\right\}. \tag{49a}$$

Solving Eq. (48a) for $n_{eh}/n_p$, we have

$$\frac{n_{eh}}{n_p} = \exp\left\{\frac{\widetilde{F}_{eh}}{T_{eh}} - \frac{\widetilde{F}_p + \widetilde{F}_{el}}{T_p + T_{el}} + \frac{T_{el}\widetilde{F}_p - T_p\widetilde{F}_{el}}{T_{eh}(T_p + T_{el})} + \frac{T_{el}(T_p + T_{eh})}{T_{eh}(T_p + T_{el})}\ln\left(1 + Z_{im}\frac{n_{im}}{n_p} - \gamma_{eh}\frac{n_{eh}}{n_p}\right)\right\}. \tag{49b}$$

Here the charge neutrality condition,

$$\frac{n_{el}}{n_p} = 1 + Z_{im}\frac{n_{im}}{n_p} - \gamma_{eh}\frac{n_{eh}}{n_p} \tag{50}$$

is used in the last term of Eq. (49b). Equation (49) is the required non-linear coupled equation for $n_{im}/n_p$ and $n_{eh}/n_p$. From Eq. (47a), the density $n_p$ is given by

$$n_p = \exp\left(\frac{\widetilde{F}_p + \widetilde{F}_{el}}{T_p + T_{el}} - 1 - \frac{T_{el}}{T_p + T_{el}}\ln\frac{n_{el}}{n_p}\right). \tag{51}$$

## C. Computational method

Estimate of the 2$^{nd}$ term of $\Omega_{\alpha\phi}$ in Eq. (40a) using the non-relativistic three-fluid model [1] suggests that it is much less than the 1$^{st}$ term $B_\phi$. Therefore, we treat the 2$^{nd}$ term as correction term including $\Omega_{eh\phi}$. The partial differential equation which must be solved is only Eq. (42) and the other equations are algebraic equations. It is assumed that the computational domain is rectangular; the poloidal magnetic flux loops are aligned along its boundary, where the poloidal magnetic flux data, $\psi_{data}$, is experimentally measured. Eddy current need not to be considered for nearly stationary plasma conditions.

Step 1. Make suitable toroidal current density model, $j_{\phi-model}$, and solve Eq. (52) to find the $0^{th}$ order poloidal magnetic flux function which satisfies the boundary condition $\psi_{data}$.

$$R\frac{\partial}{\partial R}\left(\frac{1}{R}\frac{\partial \psi}{\partial R}\right) + \frac{\partial^2 \psi}{\partial Z^2} = -R j_{\phi-model}. \tag{52}$$

Step 2. Select suitable 12 profile functions,

$F_\alpha(Y_\alpha), T_\alpha(Y_\alpha), K_\alpha(Y_\alpha)$ for $\alpha = p, im, el, eh$.

Put $Y_\alpha = \psi$, find the $0^{th}$ order densities, compute the toroidal velocities using Eq.(39), update $Y_\alpha$, and compute the current densities using Eq. (41).

Step 3. Update $\psi$, solving Eq. (42) with the boundary condition $\psi_{data}$ for prescribed $Q \equiv -R j_\phi$.

Step 4. Update $Y_\alpha$, $B_\phi$, $\Omega_{\alpha\phi}$ and $u_{\alpha\phi}$.

Step 5. Solve Eq. (49) for density ratio $n_{eh}/n_p$ iteratively for given other quantities.

Step 6. Update densities $n_\alpha$ for $\alpha = p, im, el, eh$.

Step 7. If convergence is not sufficient, return to Step 3 and iterate until solution with sufficient accuracy is found.

To solve Eq. (42) and Eq. (52) we use the second-order finite difference method with equal mesh intervals in the (R, Z) directions. To accelerate numerical computation, a successive over relation (SOR) and a progressive multi-grid scheme with three grids are combined. The $100 \times 100$ mesh numbers are used in the finest grid.

## V. EXAMPLE EQUILIBRIA

Solenoid-free ECRH-initiated and sustained ST plasmas has been studied on the EXL-50 experiment.[2] The relativistic four-fluid model is applied to compute Equilibrium #1, which approximately reproduces a set of measurements for a plasma at time=3s during shot #4851. TABLE 1 shows the measured parameters of this plasma and Fig.1 a visible image of the plasma collected by a high resolution fast camera. This plasma contains a strong energetic electron (eh) component with a significant relativistic effect and a minute boron (b) component as an impurity species, in addition to a main ion species proton (p) component and a thermal electron (el) component.

Selection of the profile functions and the reference values, especially $I_{ref}$ and $n_{ref}$ is made through trial and error and is provided in Appendix B. Fig.2 shows the computational zone, limiter front edges, and the locations of magnetic flux loops and the PF coils of EXL-50. Fig.3 shows the observed poloidal magnetic flux $\psi_{data}$ of this plasma pulse and its interpolation (solid lines) which is used as the boundary value in solving Eq. (42) and Eq. (52). Fig.4 compares the poloidal magnetic flux generated by the plasma current measured by the flux loops with the poloidal magnetic flux $\psi_{plasma}$ at the same locations computed from the plasma current of the computed equilibrium. This near-coincidence is obtained through variations of the parameter $\psi_{criteh}$ (=0.06, a normalized dimensionless value), which determines the energetic-electron-fluid boundary location, as discussed in Appendix B. TABLE 2 shows the parameters of this computed four-fluid equilibrium. Comparing with TABLE 1, the computed equilibrium deviates in the plasma current and the line-density by an error of 2.9% and 5.7%, respectively. The computed outer last closed flux surface (lcfs) radius location of $R_{lcfs\_out} = 0.786m$ is significantly smaller than that estimated based on optical data analysis $R_{lcfs} = 0.879m$[21]. Due to a large difference of $\psi_{criteh} - \psi_{lcfs}$, the outer radius of toroidal current boundary, $R_{j\phi out}$ is larger than $R_{lcfs\_out}$. The current, temperature and flow velocity of the energetic-electron-fluid are much larger than the other fluids, as shown in Fig.5. Fig.6 shows the radial force balance of each fluid on the mid-plane. The red lines, the thick blue lines and yellow lines depict the pressure gradient force, the $j_\phi B_z$ Lorentz force and the electric force, respectively. Although these forces are dominant, the centrifugal force of the proton-fluid is substantial. The force balance of the boron fluid which is shown by light-blue b-bal line is relatively large. Since magnitude of these force is $10^{-4}$ of the force acting on the energetic-electron-fluid, this error is considered acceptable. Fig.7 shows contour plots of magnetic flux, total toroidal current density, the energetic-electron-fluid current density and current density of the thermal components which are confined within the lcfs region. A relatively small amount of energetic-electron-fluid current density $J_{eh\phi}$ reaches beyond the top and bottom limiter edges, and is considered acceptable for the purpose of this paper.

      Properties of Equilibrium #2 is shown in the second column of TABLE 2 and Figs. 8-10, which aims to double the plasma current of Equilibrium #1. Since global nature of profiles of these two equilibria are similar, we neglect repeating a detailed explanation here. Equilibrium #2 contains a substantial relativistic effect as shown in Figs. 11. The energetic-electron-fluid temperature is larger than the electron rest-mass energy, $g_{ep}$ reaching up to 6.7. As a result, the centrifugal force acting on the energetic-electron-fluid increases as shown in Fig.9 (d). The computed plasma sizes and densities are also similar in value. And also the input parameters shown in TABLE 3 are the same except the CKel0 parameter, which determines the vacuum toroidal magnetic field strength. Even though, the current and temperature increase in Equilibrium #2 are clear. This method is very effective to increase current and temperature while keeping density and plasma size nearly unchanged.

TABLE 1. Parameters of EXL-50 shot #4851 at 3s.

| |
|---|
| $I_{plasma} = -50.78\ kA$ with positive $B_{\phi\_external}$ |
| $n_e l = 5.35 \times 10^{17}\ m^{-2}$ at z=0, R=0.49m. |
| Bt(at $R = 0.56m) = 0.402\ T$ |
| $T_{eh} \approx 144 \pm 30 keV$ (chord averaged value based on HXR measured from a plasma pulse under similar conditions, the peak local value is estimated to be above 200keV.) |

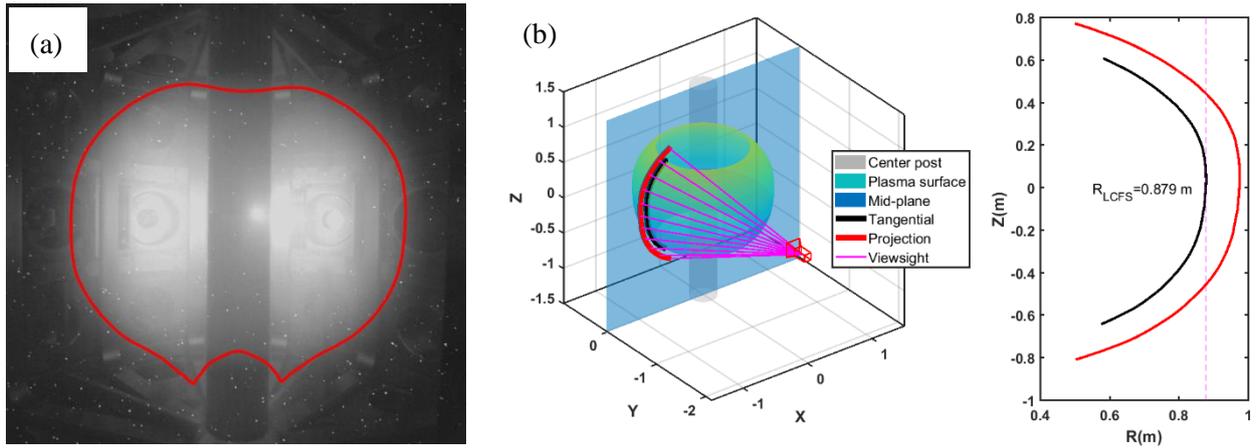

Fig. 1. (a) Plasma image of shot #4851 at 3s on the EXL-50, obtained by a high resolution fast camera (model: Phantom V1212) located at a mid-plane window. (b) An approximate reconstruction of the outer portion of an axisymmetric LCFS[21] is overlaid onto the image and is reproduced in the equilibrium computation. The red line is the projection in the vertical plane of the tangential sightline where the brightness gradient is maximum. The black line indicates the axisymmetric plasma boundary obtained via coordinate transforming. In (a), the transient random bright specs indicate the presence of strong X-ray photons, believed to have originated from a copious population of energetic electrons.

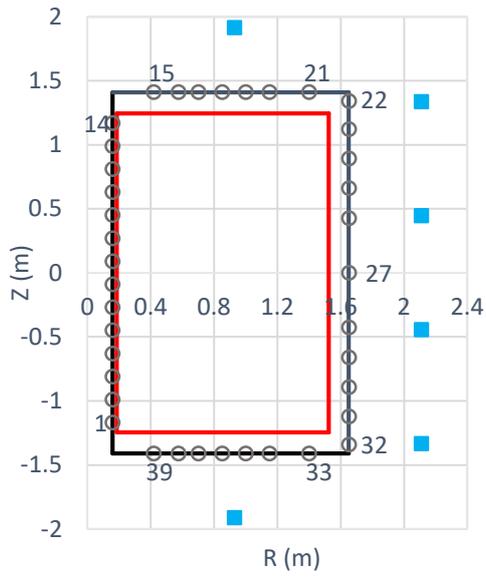

Fig. 2. Computational Zone. Black lines depict computational boundary and red lines the edge of the limiters. Circles (numbered from 1 to 39) mark the positions of magnetic flux loops and blue squares the positions of the PF coils.

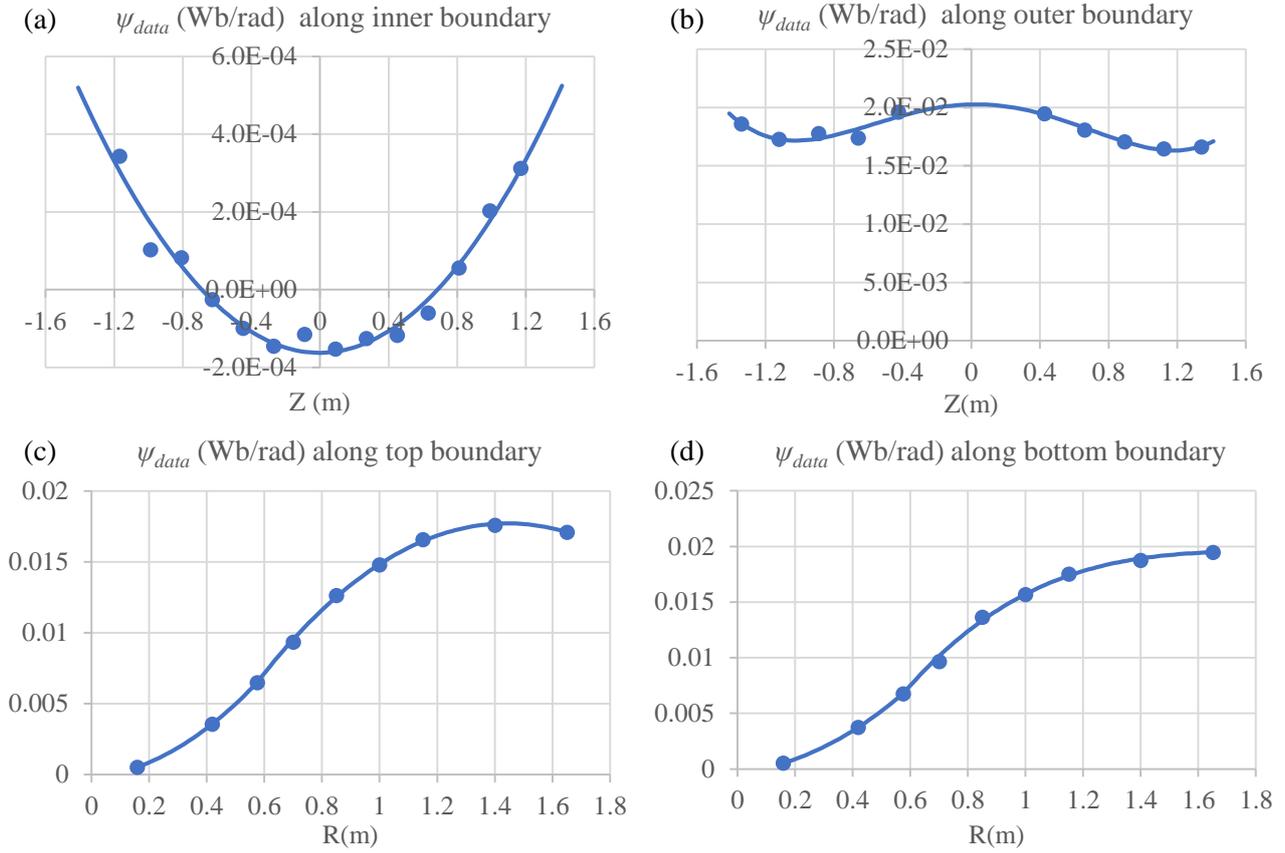

Fig. 3. Observed poloidal magnetic flux $\psi_{data}$ (Wb/rad) of EXL-50 shot #4851 at 3s (dots) and the fitted continuous values (solid lines) used in the numerical computation: (a) along inner boundary, (b) along outer boundary, (c) along top boundary and (d) along bottom boundary.

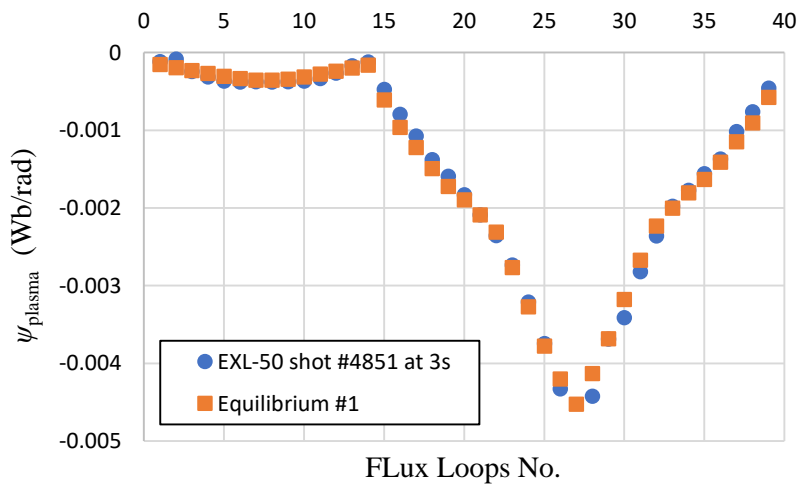

Fig. 4. Poloidal magnetic flux generated by the EXL-50 shot #4851 plasma current at 3s measured by the flux loops (dots), and the poloidal magnetic flux $\psi_{plasma}$ at the same locations computed from the plasma current of Equilibrium #1 (squares).

TABLE 2. Parameters of the computed four-fluid plasma equilibria.

| | **Equilibrium #1** (near parameters shown in Table 1 and Figs. 3-5) | **Equilibrium #2** (by doubling the boundary values of $\psi_{data}$ of Equilibrium #1) |
|---|---|---|
| Iplasma (kA) | -52.257 | -102.716 |
| $n_e l$ (m$^{-2}$) at z=0, R=0.49m | 5.66× 10$^{17}$ | 7.76 × 10$^{17}$ |
| Bt (at R=0.56m) (T) | 0.402 | 0.402 |
| Teh_max (keV) | 236.485 | 678.695 |
| Iproton (kA) | -0.650 | -0.974 |
| Iboron (kA) | -0.001 | -0.001 |
| Iel (kA) | -2.013 | -4.029 |
| Ieh (kA) | -49.562 | -97.711 |
| q(0) | 22.753 | 11.545 |
| R_mag(m) | 0.576 | 0.576 |
| R_major(m) | 0.472 | 0.472 |
| R_lcfs_out(m) | 0.786 | 0.786 |
| R_Jtout(m) | 1.218 | 1.218 |
| Tp_max (eV) | 91.497 | 242.593 |
| Tb_max (eV) | 84.559 | 235.563 |
| Tel_max (eV) | 124.900 | 353.358 |
| np_max (m$^{-3}$) | 3.96× 10$^{17}$ | 5.32× 10$^{17}$ |
| Zb*nb _max (m$^{-3}$) | 3.34× 10$^{15}$ | 5.72× 10$^{15}$ |
| nel_max (m$^{-3}$) | 3.93× 10$^{17}$ | 5.29× 10$^{17}$ |
| neh_max (m$^{-3}$) | 6.47× 10$^{15}$ | 9.13× 10$^{15}$ |
| $u_{p\phi\_max}$ (km/s) | -131.066 | -127.435 |
| $u_{b\phi\_max}$ (km/s) | -20.190 | -24.270 |
| $u_{el\phi\_max}$ (km/s) | 293.195 | 436.555 |
| $u_{eh\phi\_max}$ (km/s) | 43528.551 | 67908.283 |

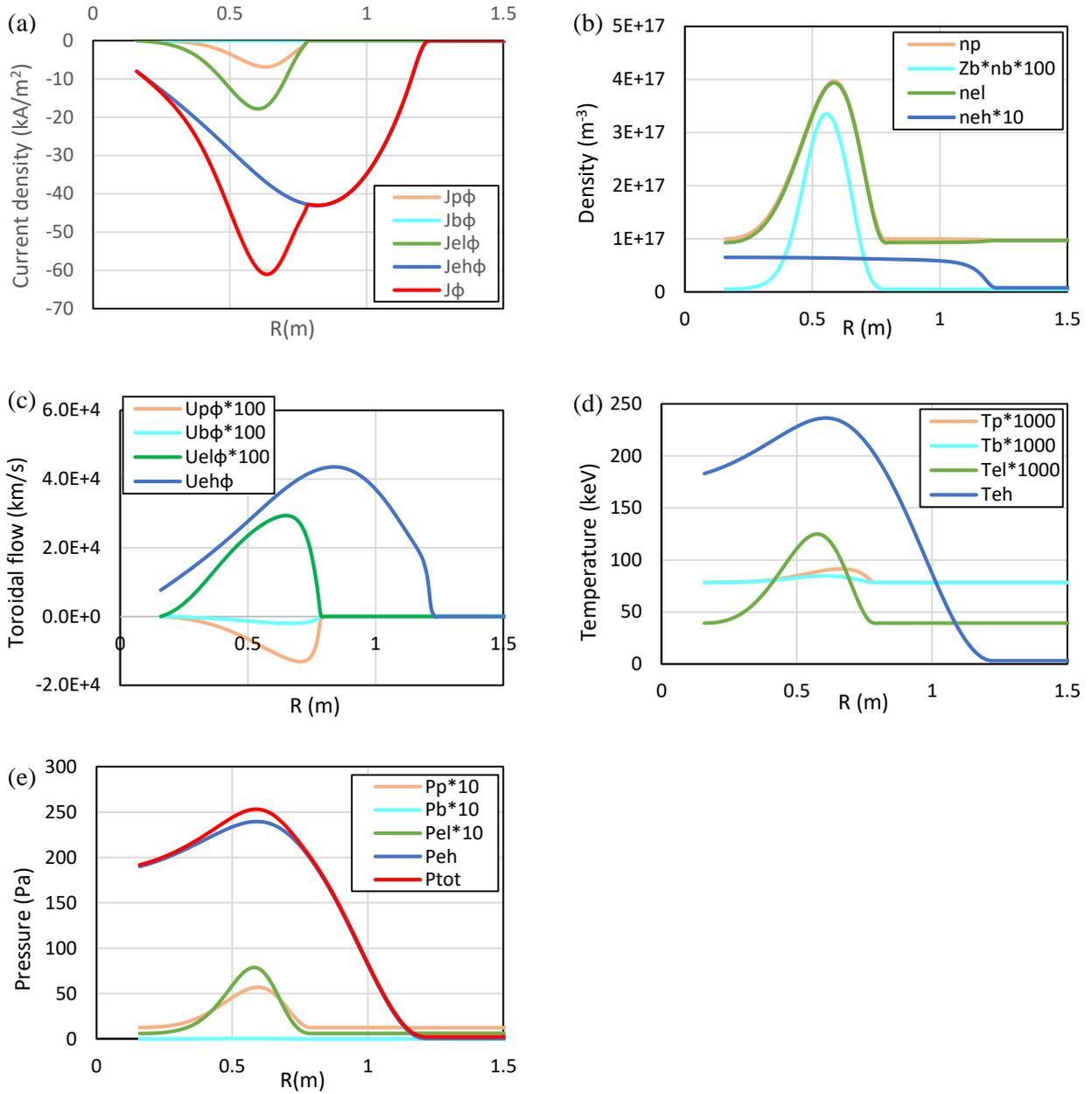

Fig. 5. Various mid-plane profiles of Equilibrium #1: (a) Toroidal current densities of proton (p), boron (b), thermal electron (el), energetic electron (eh) fluids, and their total; (b) Densities of these species; (c) Toroidal flow velocities of these species; (d) Temperatures of these species; (e) Pressure of these species and the total.

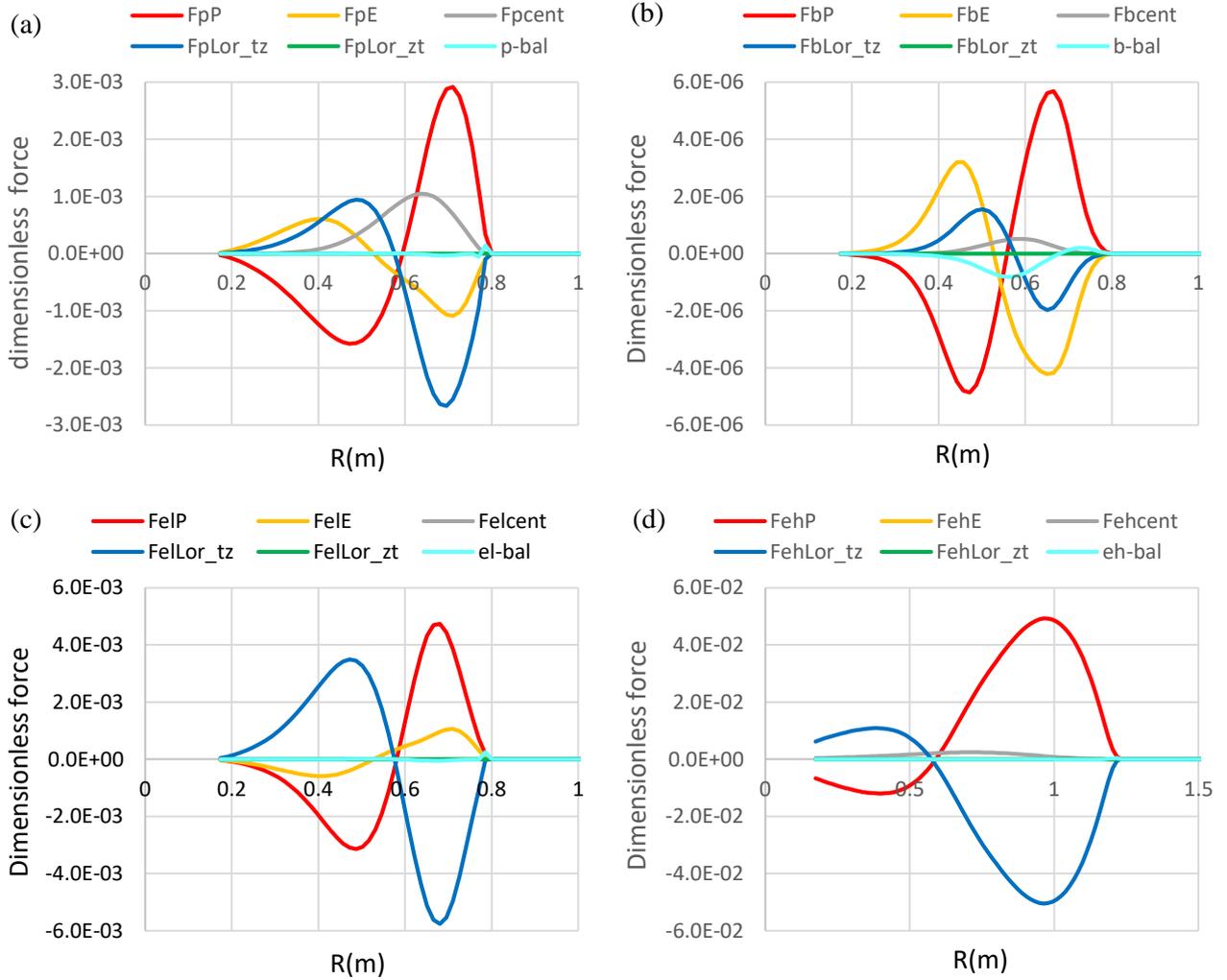

Fig.6. Radial force balance of Equilibrium #1 at the mid-plane for (a) the proton fluid, (b) the boron fluid, (c) the thermal electron fluid and (d) the energetic electron fluid. Meaning of symbols in these figures are described in Appendix C.

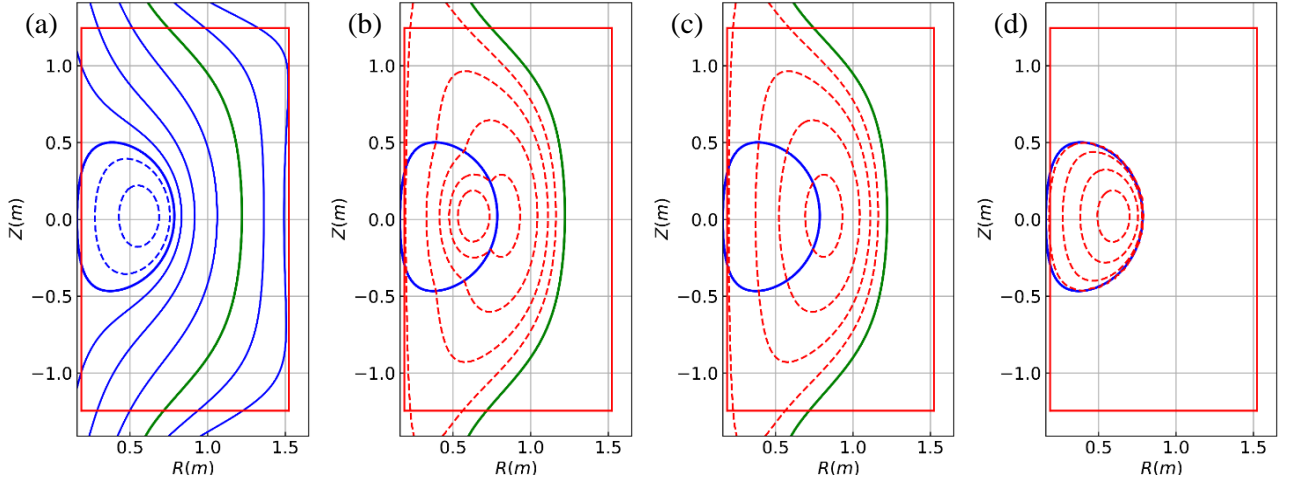

Fig.7. Various 2D contour plots of Equilibrium #1: (a) Contours of poloidal magnetic flux $\psi$, from inside to outside $\psi = $ [-0.006, -0.003, -0.001292444, 0.002, 0.01, 0.03, 0.06, 0.09, 0.12] * $\psi_{ref}$ with $\psi_{ref} = $ 0.125664 Wb/rad. Green line denotes $\psi_{criteh}$ which is 0.06 * $\psi_{ref}$. (b) Contours of total toroidal current density $J_\phi$, from inside to outside $J_\phi = $ [-0.5, -0.4, -0.3, -0.2, -0.1]* $J_{ref}$ with $J_{ref} = $100 kA/m². Blue line denoted $\psi_{lcfs}$ which is -0.001292444 *$\psi_{ref}$. (c) Contours of toroidal current density carried by the energetic electron, from inside to outside $J_{eh\phi} = $ [-0.4, -0.3, -0.2, -0.1]* $J_{ref}$ with $J_{ref} = $100 kA/m². (d) Contours of sum of the toroidal current density carried by the proton, boron and thermal electron fluid. From inside to outside $J_{p\phi} + J_{b\phi} + J_{el\phi} = $ [-0.15, -0.05, -0.01, -0.001]* $J_{ref}$ with $J_{ref} = $100 kA/m².

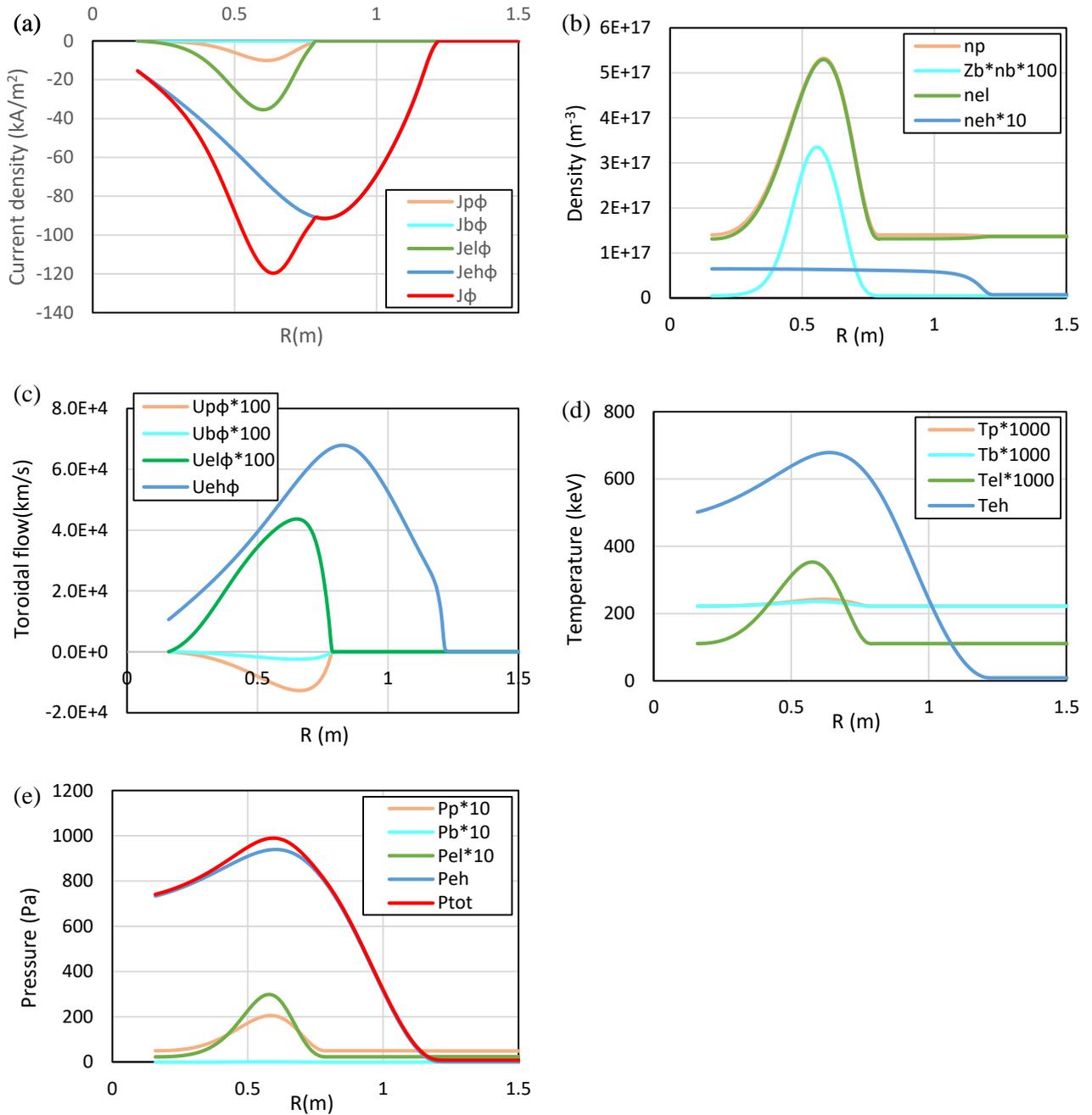

Fig.8. Various mid-plane plasma parameter profiles of Equilibrium #2: (a) Toroidal current densities of proton (p), boron (b), thermal electron (el), energetic electron (eh) fluids, and their total; (b) Densities of these species; (c) Toroidal flow velocities of these species; (d) Temperatures of these species; (e) Pressure of these species and the total.

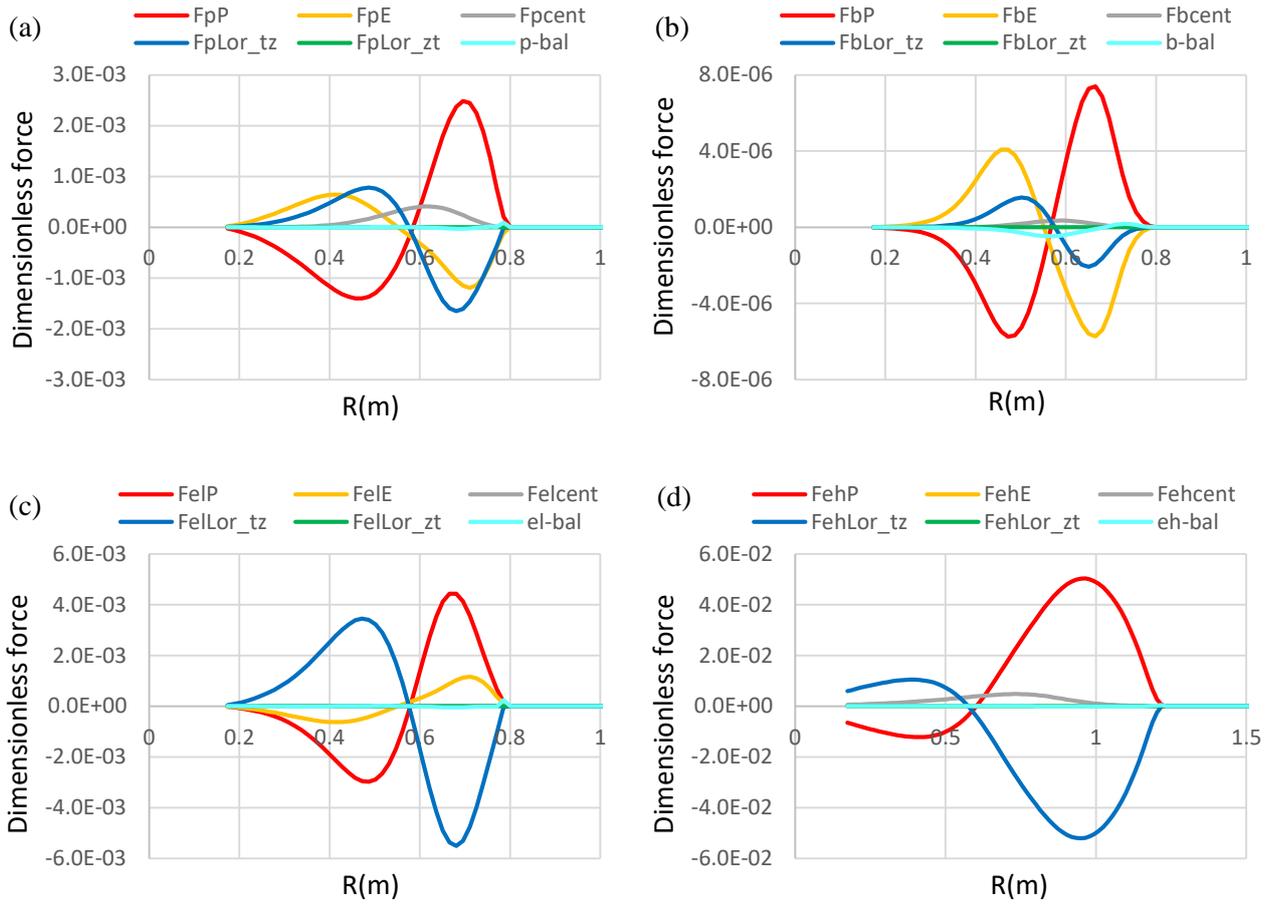

Fig.9. Radial force balance of Equilibrium #2 at the mid-plane for (a) the proton fluid, (b) the boron fluid, (c) the thermal electron fluid and (d) the energetic electron fluid. Meaning of symbols in these figures are described in Appendix C.

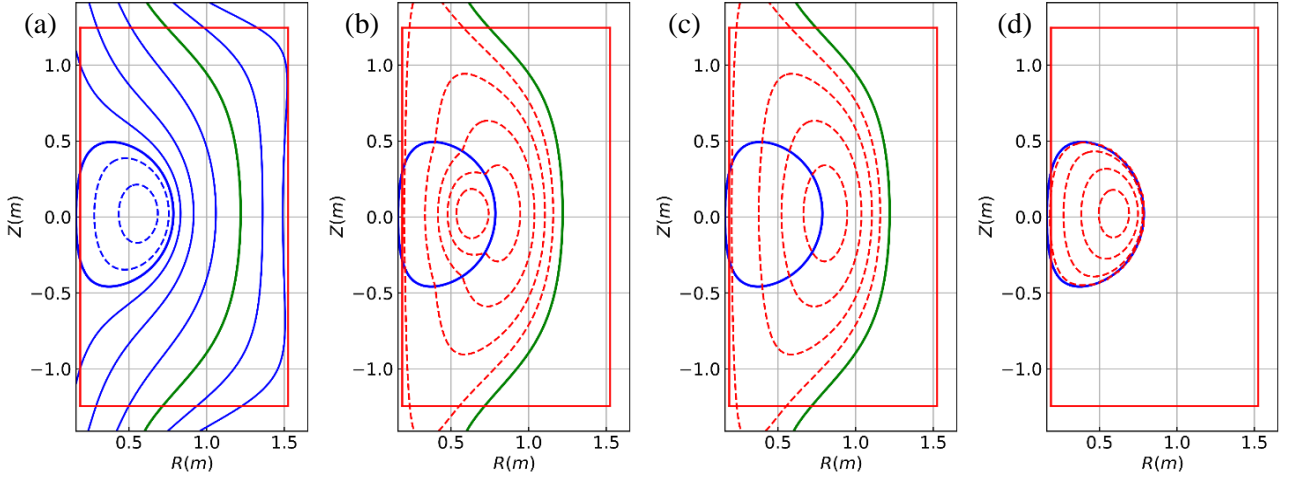

Fig.10. Various 2D contour plots of Equilibrium #2: (a) Contours of poloidal magnetic flux $\psi$, from inside to outside $\psi = $ [-0.006, -0.003, -0.001292444, 0.002, 0.01, 0.03, 0.06, 0.09, 0.12] * $\psi_{ref}$ with $\psi_{ref} = $ 0.251327 Wb/rad. Green line denotes $\psi_{criteh}$ which is 0.06 * $\psi_{ref}$. (b) Contours of total toroidal current density $J_\phi$, from inside to outside $J_\phi = $ [-0.5, -0.4, -0.3, -0.2, -0.1]* $J_{ref}$ with $J_{ref}$ =200 kA/m$^2$. Blue line denoted $\psi_{lcfs}$ which is -0.001292444 *$\psi_{ref}$. (c) Contours of toroidal current density carried by the energetic electron, from inside to outside $J_{eh\phi} = $ [-0.4, -0.3, -0.2, -0.1]* $J_{ref}$ with $J_{ref}$ =200 kA/m$^2$. (d) Contours of sum of the toroidal current density carried by the proton, boron and thermal electron fluid. From inside to outside $J_{p\phi} + J_{b\phi} + J_{el\phi} = $ [-0.15, -0.05, -0.01, -0.001]* $J_{ref}$ with $J_{ref}$ =200 kA/m$^2$.

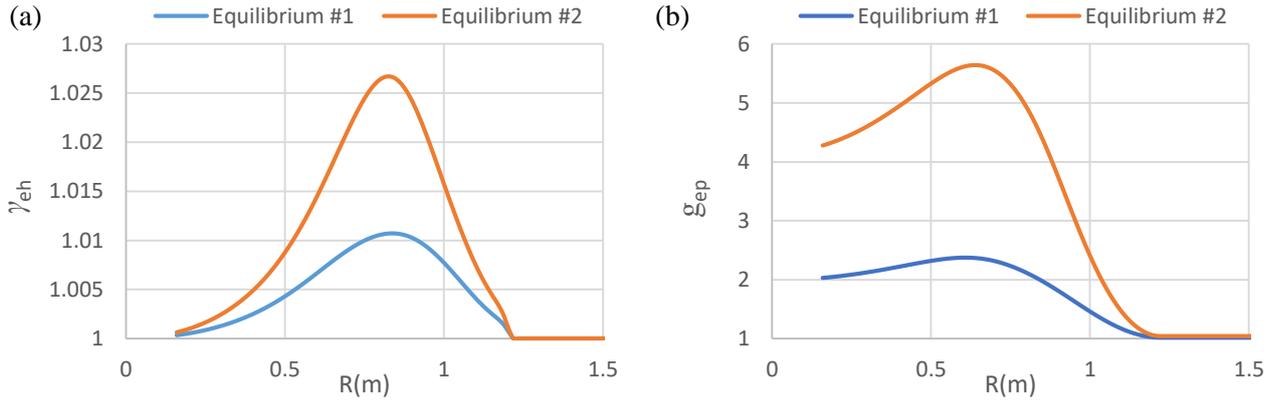

Fig.11. Comparison of the relativistic effects obtained in the two equilibria: (a) for the relativistic effects due to large macroscopic flow speed of the energetic electrons, $\gamma_{eh}(u_{eh}) = (1 - u_{eh}^2/c^2)^{-1/2}$, (b) for the relativistic effects due to large microscopic random motion of particles contained in the high-temperature energetic electron fluid, $g_{ep}(T_{eh}/m_e c^2)$, (see Eq.(3)).

## VI. SUMMARY AND DISCUSSION

We develop in this paper an axisymmetric equilibrium formulation of a relativistic multi-fluid plasma of a low aspect ratio. Based on the equations of continuity and motion for a relativistic fluid species, we derive the equation for generalized momentum $\boldsymbol{P}$, Eq. (9), which accounts for the effects of pressure gradient, electric field, relativistic macroscopic velocity through $\gamma$, and relativistic velocity of microscopic random motion through $g_{ep}(T^*)$. It is found that the z-component of generalized angular momentum, $RP_\phi$, defined in Eq. (10b), plays an important role in the axisymmetric equilibrium formulation.

The formulation is applied to a four-fluid equilibrium including two ion species such as proton and a dominant impurity ion and two electron fluid components (a low-temperature high-density thermal electron fluid and a high-temperature high-speed low-density relativistic electron fluid). As a solution for the Grad-Shafranov equation requires two profile functions, the present four-fluid model requires twelve profile functions. The derived equations for densities are consistent with the electrostatic potential.

An effective method to calculate four-fluid axisymmetric equilibrium is shown and applied to an equilibrium observed in the EXL-50 experiment (Equilibrium #1), which contains boron as an impurity and energetic electrons that carry a large fraction of the plasma current that extends over the open field region. As a second equilibrium (Equilibrium #2), which doubles the plasma current of Equilibrium #1 while maintaining the toroidal field strength (by adjusting the input coefficient CKel0), the calculated equilibrium central temperature and density of the energetic-electron component (eh-fluid) are $T_{eh}$ =679 keV and $n_{eh}$ =9.13E+15 $m^{-3}$, while those of the thermal-electron component (el-fluid) are $T_{el}$ = 355eV and $n_{el}$ = 5.3E+17 $m^{-3}$.

The computed thermal and the energetic electron fluids can have quite different parameters, the latter fluid containing a clear large relativistic effect. By comparing the detailed features of Equilibria #1 and #2, one obtains from Fig. 11 that the relativistic effects due to large macroscopic flow speed of the energetic electrons, $\gamma_{eh}(u_{eh})$, are relatively modest (increasing from a central value of 1.01 to nearly 1.03), and the relativistic effects due to large microscopic random motion of particles contained in the high-temperature energetic electron fluid, $g_{ep}(T_{eh}/m_e c^2)$, are pronounced (increasing from a central value of 2.3 to 5.6).

As more experimental results emerge, Equilibrium #2 will be updated according to measurements. Effects of the energetic electron energy tail on the equilibrium is not included in the present model. It may become of interest if an energy tail were to become significant in future EXL-50 plasmas.

## ACKNOWLEDGMENTS

The authors acknowledge the permission by the EXL-50 team to quote the measured parameters of a plasma for comparison with the equilibria computed in this paper. In particular, we acknowledge permission by Guo Dong to utilize poloidal flux related data, Dr. Deng Bihe and Li

Songjian the line-density data, Dr. Liu Bing the fast camera data, and Dr. Shi Yuejiang the HXR data. We further thank Guo Dong for discussion on an improved technique for estimating the outer location of the last closed flux surface of a plasma produced in the EXL-50. One of the authors AI would like to thank Dr. Shi Yuejiang for discussion on X-ray spectroscopy of energetic electrons.

# APPENDIX A: THERMODYNAMIC PROPERTY OF RELATIVISTIC IDEAL FLUID[8]

In this Appendix, let us consider energy density $\epsilon$ and pressure $p$ of an ideal fluid. The thermal equilibrium can be described only in the proper frame in which the fluid volume element concerned is at rest. Also note that as a particle energy includes a rest-mass energy in relativistic theory, the internal energy density $\epsilon$ also includes the rest-mass energy.

The distribution function $f$ of thermal equilibrium is given by

$$f(\boldsymbol{p}) = \frac{g_{spin}}{h^3} \frac{1}{e^{(E-\mu)/T} \pm 1} , \qquad (A.1)$$

where $\boldsymbol{p}$ is momentum of a particle, $g_{spin}$ is degree of spin freedom, $h$ is Plank's constant, $E$ is energy of a particle, $\mu$ is chemical potential and $T$ is temperature in energy unit. The – sign is for Bose particle and the + sign for Fermi particle in the denominator. For dilute gas such as plasma fluid considered here, the $\pm 1$ term can be neglected.

$$f(\boldsymbol{p}) = \frac{g_{spin}}{h^3} e^{(\mu-E)/T}. \qquad (A.2)$$

Since the proper frame is local and instantaneous, the distribution function $f$ depends on time $t$ and coordinate $\boldsymbol{x}$ (local and temporal thermal equilibrium), we neglect these two dependencies for simplicity.

Let us calculate the number density $n$,

$$n = \int d^3\boldsymbol{p} f(\boldsymbol{p}) \qquad (A.3)$$

Since a particle's momentum and energy are given by $\boldsymbol{p} = m\gamma\boldsymbol{v}$ where $\gamma = (1 - v^2/c^2)^{-1/2}$ and $E = mc^2\gamma$, we have

$$\frac{E}{mc^2} = \sqrt{1 + \left(\frac{p}{mc}\right)^2}. \qquad (A.4)$$

Putting $x \equiv \frac{p}{mc}$, Eq. (A.3) is written as,

$$n = 4\pi g_{spin} e^{\mu/T} \left(\frac{mc}{h}\right)^3 \int_0^\infty dx\, x^2 e^{\frac{-1}{T^*}\sqrt{x^2+1}} . \qquad (A.5)$$

Again putting $t \equiv \sqrt{x^2 + 1}$ and using $de^{-zt}/dz = -t\, e^{-zt}$, the integral of the above equation is written as,

$$\int_0^\infty dx\, x^2 e^{\frac{-1}{T^*}\sqrt{x^2+1}} = -\frac{d}{dz}\int_1^\infty dt\, e^{-zt}\sqrt{t^2 - 1} \qquad (A.6)$$

where $z \equiv 1/T^*$.

Integral representation of the second kind modified Bessel function $K_n(z)$ is given by

$$K_n(z) = \frac{\sqrt{\pi}\left(\frac{z}{2}\right)^n}{\Gamma\left(n+\frac{1}{2}\right)} \int_1^\infty dt e^{-zt}(t^2-1)^{n-\frac{1}{2}} \tag{A.7}$$

where $\Gamma$ is the Gamma function and $K_n(z)$ satisfies the following recurrence relations.

$$K_{n-1}(z) + K_{n+1}(z) = -2K_n'(z). \tag{A.8a}$$

$$K_{n-1}(z) - K_{n+1}(z) = -\frac{2n}{z}K_n(z). \tag{A.8b}$$

Applying Eq. (A.7) and Eq. (A.8) to Eq. (A.6), the result is given by

$$\int_0^\infty dx x^2 e^{\frac{-1}{T^*}\sqrt{x^2+1}} = -\frac{d}{dz}\left(\frac{K_1(z)}{z}\right) = \frac{K_2(z)}{z} \tag{A.9}$$

Therefore the number density $n$ of Eq. (A.5) is written as

$$n = 4\pi g_{spin} e^{\frac{\mu}{T}}\left(\frac{mc}{h}\right)^3 T^* K_2(1/T^*) \text{ where } T^* \equiv \frac{T}{mc^2}. \tag{A.10}$$

It should be understood that this equation determines the chemical potential $\mu$ for given temperature $T$ and density $n$. Using Eq. (A.10), the distribution function $f$ is written as

$$f(\boldsymbol{p}) = \frac{n}{4\pi(mc)^3 T^* K_2(1/T^*)} e^{-E/T}. \tag{A.11}$$

Next, we calculate the energy density $\epsilon$,

$$\epsilon = \int d^3\boldsymbol{p} E f(\boldsymbol{p}) = \frac{nmc^2}{T^* K_2(1/T^*)} \int_0^\infty dx x^2 \sqrt{x^2+1}\, e^{\frac{-1}{T^*}\sqrt{x^2+1}}, \tag{A.12}$$

where Eq. (A.4) and $x \equiv \frac{p}{mc}$ are used. Putting $t \equiv \sqrt{x^2+1}$ and $z \equiv 1/T^*$ and using $d^2 e^{-zt}/dz^2 = t^2 e^{-zt}$, the above integral is given by

$$\int_0^\infty dx x^2\sqrt{x^2+1}\, e^{\frac{-1}{T^*}\sqrt{x^2+1}} = \frac{d^2}{dz^2}\left(\frac{K_1(z)}{z}\right) = \frac{3K_3(z)+K_1(z)}{4z}$$

where the recurrence relations (A.8) are used. Substituting the above into Eq. (A.12) gives

$$\epsilon = nmc^2 \frac{3K_3(1/T^*)+K_1(1/T^*)}{4K_2(1/T^*)}. \tag{A.13}$$

Finally, let us consider the pressure $p$. Suppose an ideal gas such as plasma is confined in a small cubic box with volume $l^3$. When a particle with momentum $p_x$ hits the wall, momentum change $\Delta p = 2p_x$ in time interval $\Delta t = 2l/v_x$. Pressure due to single particle acting on the y-z plane is given by $\frac{\Delta p}{\Delta t l^2} = \frac{p_x v_x}{l^3}$. Thus, pressure $p$ is given by

$$p = nl^3 \langle\frac{p_x v_x}{l^3}\rangle = \int d^3\boldsymbol{p}\, p_x v_x f(\boldsymbol{p}) = \int d^3\boldsymbol{p} \frac{1}{3}\boldsymbol{p}\cdot\boldsymbol{v} f(\boldsymbol{p})$$

Since $\bm{p} \cdot \bm{v} = mc^2 \frac{x^2}{\sqrt{x^2+1}}$ where $x \equiv \frac{p}{mc}$, substituting Eq. (A.11) into the above,

$$p = \frac{nmc^2}{3} \frac{z}{K_2(z)} \int_0^\infty dx \frac{x^4}{\sqrt{1+x^2}} e^{-z\sqrt{1+x^2}} \text{ where } z \equiv 1/T^*.$$

Changing variable $t \equiv \sqrt{x^2 + 1}$, the above integral becomes $3K_2(z)/z^2$. As a result, the pressure $p$ is written as

$$p = nmc^2 T^* = nT \tag{A.14}$$

From Eq. (A.13) and Eq. (A.14) we have useful relation (A.15), defining the relativistic energy density ratio over that of the fluid rest-mass energy density, $g_{ep}(T^*)$.

$$\frac{\epsilon+p}{nmc^2} = \frac{K_3(1/T^*)}{K_2(1/T^*)} \equiv g_{ep}(T^*) \tag{A.15}$$

where we used relation $K_1(1/T^*) - K_3(1/T^*) = -4T^* K_2(1/T^*)$.

Let us consider $\frac{K_3(1/T^*)}{K_2(1/T^*)}$ for $T^* \ll 1$ non-relativistic limit. Changing variable, $x = t - 1$ in Eq. (A.7),

$$\frac{K_3(z)}{K_2(z)} = \frac{z}{5} \frac{I_3(z)}{I_2(z)} \text{ where } I_n(z) \equiv \int_0^\infty dx e^{-zx}(x(x+2))^{n-\frac{1}{2}}. \tag{A.16}$$

For $z \gg 1$ limit, i.e. for $T^* \ll 1$ non-relativistic limit,

$$I_n(z) = 2^{n-\frac{1}{2}} \left( Y_n(z) + \tfrac{1}{2}(n-\tfrac{1}{2}) Y_{n+1}(z) \right)$$

where $Y_n(z) = \int_0^\infty dx e^{-zx} x^{n-\frac{1}{2}} = (-1)^n \sqrt{\pi} \frac{d^n z^{-\frac{1}{2}}}{dz^n}$. Using this, we have

$$\frac{I_3(z)}{I_2(z)} = \frac{5}{z}(1 + \frac{5}{2z}) \text{ and } \frac{K_3(z)}{K_2(z)} = 1 + \frac{5}{2z} \text{ for } z \gg 1.$$

Finally we have for $T^* \ll 1$ non-relativistic limit, $\frac{K_3(1/T^*)}{K_2(1/T^*)} = 1 + \frac{5}{2}T^*$. \quad (A.17)

Next, consider relativistic limit, $T^* \gg 1$ or $z \ll 1$. In this limit the integral $I_n(z)$ is approximated as,

$$I_n(z) = \int_0^\infty dx\, e^{-zx} x^{2n-1}. \text{ As a result, } \frac{K_3(z)}{K_2(z)} \to \frac{4}{z} \text{ i.e. } \frac{K_3(1/T^*)}{K_2(1/T^*)} \to 4T^*. \tag{A.18}$$

For convenience, the following approximation is often used.[9-11]

$$\frac{\epsilon+p}{nmc^2} = 1 + \frac{\Gamma}{\Gamma-1} T^* \text{ where } \Gamma = \frac{5}{3} \text{ for non-relativistic limit and } \Gamma = \frac{4}{3} \text{ for relativistic limit.}$$

In a relativistic two-fluid model of electrons and positrons, $\Gamma = \frac{4}{3}$ is used in Ref.11.

Figure 12 shows computation result. Red circles depict computational results of $\frac{K_3(1/T^*)}{K_2(1/T^*)}$. Blue line depicts interpolation result which is given by for $0 \leq T^* = \frac{T}{mc^2} \leq 2$, $g_{ep}(T^*) = \frac{K_3(1/T^*)}{K_2(1/T^*)} = 1 + \frac{5}{2}T^* + 1.1495T^{*2} - 0.3025T^{*3}$. (A.19)

Since $mc^2 \cong 0.5MeV$ for electron fluid, $g_{ep}$ which shows relativistic enhancement due to particle's random motion becomes ~4 at T=0.5MeV. Note that incline of the curve is almost 4 for the interval $1.0 \leq T^* \leq 2.0$.

Figure 13 shows contour plot of function $\gamma g_{ep}(T^*)$ which represents relativistic enhancement of difference $(Y_{eh} - \psi)$. Since the surface function $Y$ defined by Eq. (20) is equal to $RP_\phi/q$ where $RP_\phi$ is the z-component of generalized angular momentum defined by Eq. (10b), $\gamma g_{ep}$ depicts relativistic enhancement of momentum due to macroscopic motion and microscopic random motion. Equation (37d) shows that surface function $Y_{eh}$ depends on $\gamma g_{ep}$. It is useful to know how large this quantity is.

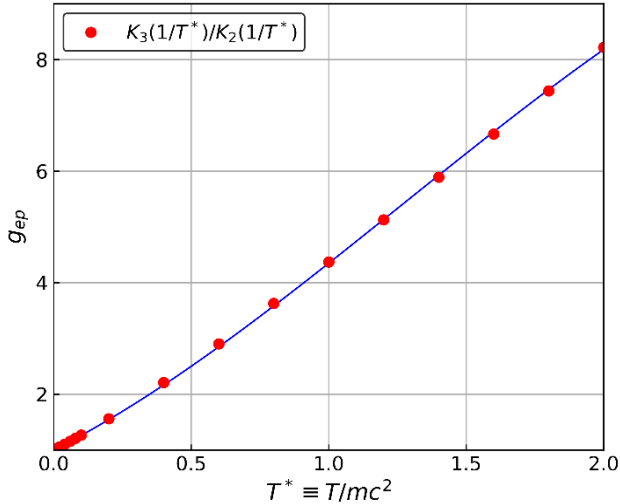

**Fig. 12** Red circles depict direct computational results of the second kind modified Bessel function, $\frac{K_3(1/T^*)}{K_2(1/T^*)}$. Blue line shows interpolation result of Eq. (A.18). This expression may be useful for computation.

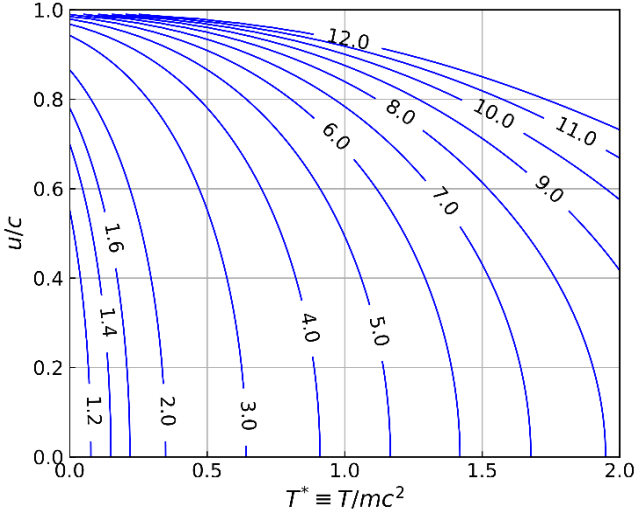

**Fig. 13** Contour plot of $\gamma g_{ep}(T^*)$ where $\gamma = (1 - u^2/c^2)^{-1/2}$ and $T^* = T/mc^2$.

## APPENDIX B: PROFILE FUNCTIONS

Various profile function formats were tested and the followings were found to work well.

Define generic function $g_\alpha(x)$ for the $\alpha$ species as

For $x \geq 0$, $g_\alpha(x; c_\alpha) = c_\alpha x^2$ and $g'_\alpha(x; c_\alpha) = 2c_\alpha x$

For $x < 0$, $g_\alpha(x; c_\alpha) = 0$ and $g'_\alpha(x; c_\alpha) = 0$

The profile function $F_\alpha(Y_\alpha)$ of the $\alpha$ species is given by

For $Y_\alpha - \psi_{crit\alpha} \leq 0$,

$$F_\alpha(Y_\alpha) = C_{F\alpha 0} + C_{F\alpha 1} g_\alpha(-(Y_\alpha - \psi_{crit\alpha}); c_\alpha)$$

$$F'_\alpha(Y_\alpha) = -C_{F\alpha 1} g'_\alpha(-(Y_\alpha - \psi_{crit\alpha}); c_\alpha)$$

For $Y_\alpha - \psi_{crit\alpha} > 0$,

$$F_\alpha(Y_\alpha) = C_{F\alpha 0} \text{ and } F'_\alpha(Y_\alpha) = 0$$

The profile function $T_\alpha(Y_\alpha)$ is defined by the above same format using different coefficients $C_{T\alpha 0}$ and $C_{T\alpha 1}$.

The generic function of the profile function $K_\alpha(Y_\alpha)$ is defined as

For $x \geq 0$, $g_K(x; c_K) = c_K x^3$, $g'_K(x; c_K) = 3c_K x^2$, and $g''_K(x; c_K) = 6c_K x$.

For $x < 0$, $g_K(x; c_K) = 0$, $g'_K(x; c_K) = 0$ and $g''_K(x; c_K) = 0$.

The profile function $K_\alpha(Y_\alpha)$ of the $\alpha$ species is given by

For $Y_\alpha - \psi_{crit\alpha} \leq 0$,

$$K_\alpha(Y_\alpha) = C_{K\alpha 0} + C_{K\alpha 1} g_K(-(Y_\alpha - \psi_{crit\alpha}); c_K).$$

$$K'_\alpha(Y_\alpha) = -C_{K\alpha 1} g'_K(-(Y_\alpha - \psi_{crit\alpha}); c_K).$$

$$K''_\alpha(Y_\alpha) = C_{K\alpha 1} g''_K(-(Y_\alpha - \psi_{crit\alpha}); c_K).$$

For $Y_\alpha - \psi_{crit\alpha} > 0$,

$$K_\alpha(Y_\alpha) = C_{K\alpha 0}, \quad K'_\alpha(Y_\alpha) = 0 \text{ and } K''_\alpha(Y_\alpha) = 0.$$

Note that $\psi_{critp} = \psi_{critb} = \psi_{critel} = \psi_{lcfs}$ and $\psi_{criteh} = 0.06$ are chosen in real computation.

TABLE 3. Input parameters and coefficients for Equilibria #1 and #2.

|  | **Equilibrium #1** | **Equilibrium #2** |
|---|---|---|
| Lref (m) | 1 | 1 |
| Iref (kA) | 100 | 200 |
| Nref (m$^{-3}$) | $1 \times 10^{18}$ | $\sqrt{2} \times 10^{18}$ |
| R$_{jt}$ | 0.6 | 0.6 |
| aR | 0.5 | 0.5 |
| aZ | 0.7 | 0.7 |
| C2 | 0.16 | 0.16 |
| C3 | 2 | 2 |
| $\psi_{lcfs}$ | -0.001292444 | -0.001292444 |
| $\psi_{criteh}$ | 0.06 | 0.06 |
| gelC1 | 29.031 | 30.40745 |
| gehC1 | 0.2029 | 0.1961 |
| gpC1 | 11 | 11 |
| gbC1 | 11 | 11 |
| gKC1 | 0.5 | 0.5 |
| CFp0 | -0.001 | -0.001 |
| CFp1 | 1 | 1 |
| CTp0 | 0.001 | 0.001 |
| CTp1 | 0.1 | 0.1 |
| CKp0 | 0 | 0 |
| CKp1 | -0.0001 | -0.0001 |
| CFel0 | -0.001 | -0.001 |
| CFel1 | 1.2 | 1.2 |
| CTel0 | 0.0005 | 0.0005 |
| CTel1 | 0.9 | 0.9 |
| CKel0 | 1.7922 | 0.8961 |
| CKel1 | 0.00001 | 0.00001 |
| CFeh0 | -0.25 | -0.25 |
| CFeh1 | -12000 | -12000 |
| CTeh0 | 0.04 | 0.04 |
| CTeh1 | 3000 | 3000 |
| CKeh0 | 0 | 0 |
| CKeh1 | -0.001 | -0.001 |
| CFb0 | -0.009 | -0.009 |
| CFb1 | 0.6 | 0.6 |
| CTb0 | 0.001 | 0.001 |
| CTb1 | 0.1 | 0.1 |
| CKb0 | 0 | 0 |
| CKb1 | -0.00001 | -0.00001 |

Note that $R_{jt}$, aR, aZ, C2 and C3 are the parameters to determine the toroidal current density model $j_{\phi-model}$. See Eq. (52). Various toroidal current density models were tested and the following one was used in this paper:

$$j_{\phi-model} = -C_2 R Exp\left\{C_3\left(1 - \left(\frac{R-R_{jt}}{aR}\right)^2 - \left(\frac{Z}{aZ}\right)^2\right)\right\}.$$

## APPENDIX C: RADIAL FORCE BALANCE AT MID-PLANE

Here boron is adopted as impurity ion.

For the proton fluid,

$$-\frac{\partial P_p}{\partial R} - n_p \frac{\partial V_E}{\partial R} + j_{p\phi}B_z - j_{pz}B_\phi + \frac{n_p u_{p\phi}^2}{R} = 0 \qquad (C.1)$$

For the boron fluid,

$$-\frac{\partial P_b}{\partial R} - Z_b n_b \frac{\partial V_E}{\partial R} + j_{b\phi}B_z - j_{bz}B_\phi + \frac{m_b}{m_p}\frac{n_b u_{b\phi}^2}{R} = 0 \qquad (C.2)$$

For the el-electron fluid,

$$-\frac{\partial P_{el}}{\partial R} + n_{el} \frac{\partial V_E}{\partial R} + j_{el\phi}B_z - j_{elz}B_\phi + \frac{m_e}{m_p}\frac{n_{el} u_{el\phi}^2}{R} = 0 \qquad (C.3)$$

For the eh-electron fluid,

$$-\frac{\partial P_{eh}}{\partial R} + \gamma_{eh} n_{eh} \frac{\partial V_E}{\partial R} + j_{eh\phi}B_z - j_{ehz}B_\phi + \frac{m_e}{m_p}\frac{n_{eh} u_{eh\phi}^2}{R}\gamma_{eh}^2 g_{ep}(T_{eh}^*) = 0 \qquad (C.4)$$

Note that $F_\alpha P$, $F_\alpha E$, $F_\alpha Lor\_tz$, $F_\alpha Lor\_zt$, $F_{\alpha cent}$ and $\alpha\_bal$ for $\alpha = p, b, el$ and $eh$ in Figs. 6 and 9 depict the 1$^{st}$, 2$^{nd}$, 3$^{rd}$, 4$^{th}$, 5$^{th}$ and sum of these five terms in Eq. (C.1) to Eq.(C.4).

**DATA AVAILABILITY STATEMENT**

The data that support the findings of this study are available from the corresponding author upon reasonable request.